\documentclass[aps,prl,twocolumn,showpacs,preprintnumbers,amsmath,amssymb]{revtex4}
\usepackage{graphicx}
\usepackage{dcolumn}
\usepackage{bm}

\begin{document}
\flushbottom

\title{Formation of Atomic Carbon Chains from Graphene Nanoribbons}

\author{Edwin Hobi Jr.$^1$}
\email{edwin@if.usp.br}
\author{Renato B. Pontes$^1$}
\author{A. Fazzio$^{1,2}$}
\author{Ant\^onio J. R. da Silva$^{1,3}$}
\email{ajrsilva@if.usp.br}

\affiliation{
$^1$Instituto de F\'{\i}sica, Universidade de S\~ao Paulo, CP 66318, 05315-970,
S\~ao Paulo, SP, Brazil\\
$^2$Centro de Ci\^ encias Naturais e Humanas, Universidade Federal do ABC, Santo Andr\'e,
S\~ao Paulo,SP, Brazil\\
$^3$Laborat\'orio Nacional de Luz S\'{\i}ncrotron, Campinas, SP, Brazil}

\date{\today}

\begin{abstract}
The formation of one-dimensional carbon chains from graphene nanoribbons is investigated using {\it ab initio} molecular dynamics. We show under what conditions it is possible to obtain a linear atomic chain via pulling of the graphene nanoribbons. The presence of dimers composed of two-coordinated carbon atoms at the edge of the ribbons is necessary for the formation of the linear chains, otherwise there is simply the full rupture of the structure. The presence of Stone-Wales defects close to these dimers may lead to the formation of longer chains. The local atomic configuration of the suspended atoms indicates the formation of single and triple bonds, which is a characteristic of polyynes.
\end{abstract}


\maketitle

Nanoelectronics continuously searches for low dimensional systems which can be used either as nanocontacts or nanoconductors. Metallic nanowires, for example, are widely studied because they can present quantum conductance and the capacity to produce atomic chains\cite{agrait03,fios-nosso}.

Carbon based-systems, such as carbon nanotubes\cite{iijima} and more recently graphene and its derivatives\cite{castro-neto-grafeno}, are another class of materials which have attracted strong interest. They present interesting mechanical and electronic properties with great potential for applications in nanodevices. Useful properties, such as stability, flexibility, charge carriers linear dispersion and high mobility, spin injection with long relaxation times and correlation lengths could lead to applications in spintronics.

The possibility to join in the same system the features of one-dimensional wires and the properties of carbon based materials is a very exciting one. The electronic and transport properties of one dimensional carbon systems have already been studied by some groups\cite{chain2,chain3,chain4,chain5,chain6,chain7,chain8}. However, the lack of reliable and effective ways to produce 1D carbon chains have limited the studies of these systems. Recently, two groups\cite{jin09,chuvilin} employed a technique similar to the one used for the fabrication of metallic quantum wires\cite{fios-au} to obtain experimentally stable and rigid carbon atomic chains, which brought new attention to this subject and opened up a new avenue in the investigation of one-dimensional carbon-based systems.

Even though this recent experimental work obtained these chains via removal of carbons atoms using the electron beam, one can envisage a situation where these chains could be obtained by stretching graphene nanoribbons\cite{muller1-92,peng}. In the present work we theoretically address this question. We perform  room temperature {\it ab initio} molecular dynamics (AIMD)\cite{hobi08} to investigate the formation of linear atomic carbon chains from graphene nanoribbons. We elucidate the mechanism of formation of these chains and show under what conditions it is possible to form these wires.

\begin{figure}[!ht]
\includegraphics[width=8.5cm]{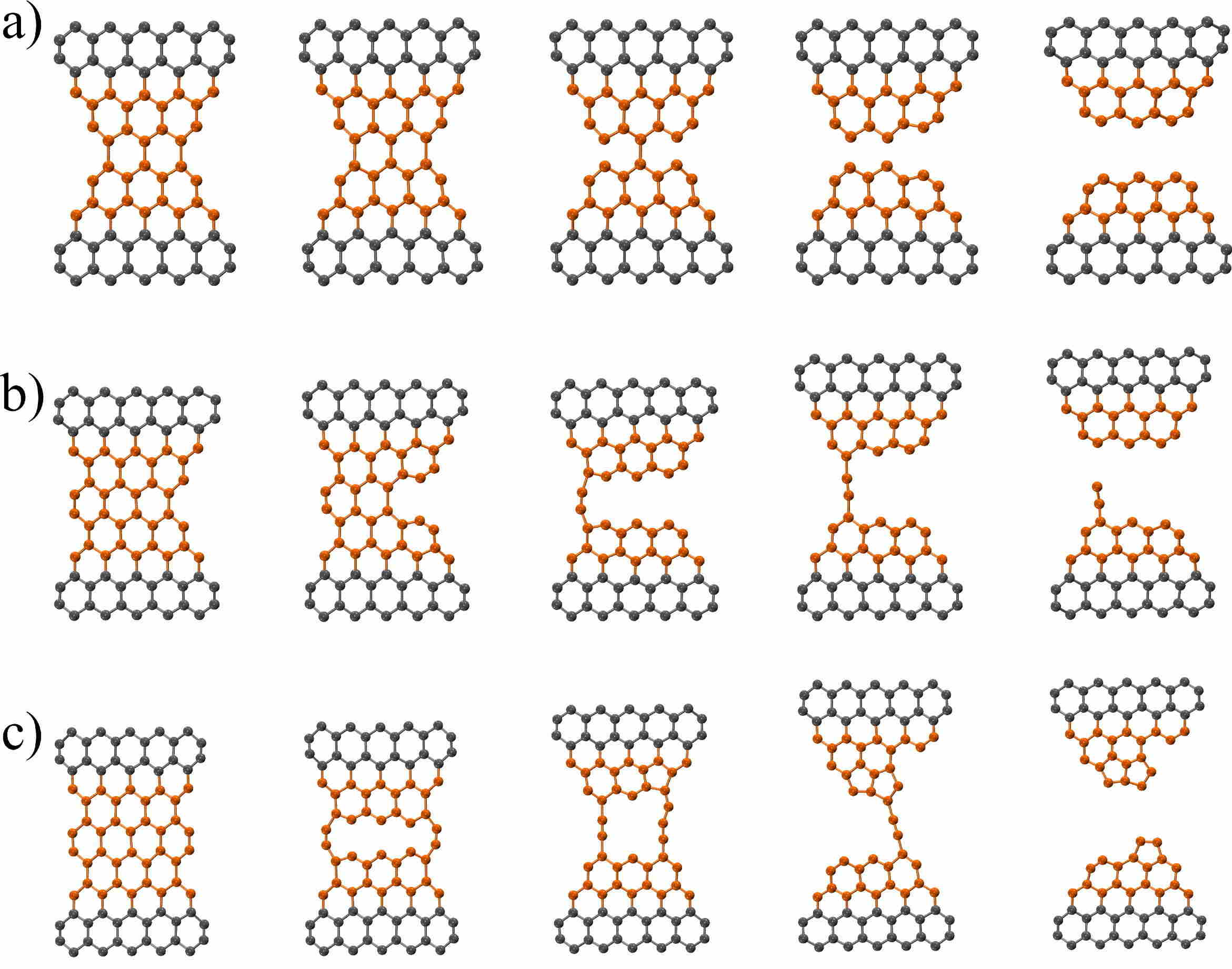}
\caption{Five representative geometries along the molecular dynamics simulations for structures (a) N1; (b) N2; and (c) N3. In (b) and (c), it is possible to see the transition from a graphene ribbon to a single carbon chain.}
\label{fig1}
\end{figure}

We study graphene nanoribbons with a neck, as shown in Fig. 1. The electronic structure and forces were obtained via {\it ab initio} total energy density functional theory\cite{dft} calculations\cite{metodo}.
We have investigated three types of necks (Fig.\ref{fig1}). In one of them the neck has only three-coordinated carbon atoms (Fig.\ref{fig1}(a)), and this will be labeled from now on N1. In the other two systems there are additional carbon dimers at the neck, which have two-coordinated carbon atoms. In one of them only one dimer is added to one of the sides (Fig.\ref{fig1}(b)), and in the other two dimers are symmetrically added to the two sides of the neck (Fig.\ref{fig1}(c)). These systems will be labeled N2 and N3, respectively. In Fig.\ref{fig1} we present five representative geometries along the molecular dynamics for each one of these systems (see also movies S1-S3\cite{movies}). As can be seen, N1 did not form any one-dimensional carbon chain whereas N2 and N3 formed one and two carbon chains, respectively. Thus, the important conclusion is that it seems necessary to have dimers of two-coordinated carbon atoms in order to form one-dimensional chains.

In more detail, the system N1 evolved in such a way that the external carbon bonds in the neck broke almost simultaneously, followed by a very rapid rupture of the central remaining bond. For N2, the rupture started at the opposite side from the two-coordinated carbons dimer, and as a zipper mechanism the other three-coordinated carbon bonds subsequently broke, creating a one-dimensional carbon chain. Upon continuous stretching this chain eventually broke. Finally, for N3, the rupture initiated at the middle of the neck, again at the three-coordinated carbon bonds. Two one-dimensional chains were formed. Upon further stretching, one of the sides presented a reconstruction where a hexagon was transformed into a pentagon with the increase of the length of the chain attached to this side. As a consequence, the continuous pulling led to the rupture of the smaller chain, which upon retraction towards the bulk formed another pentagon. Further stretching caused yet another atomic rearrangement, with the first pentagon being broken and reconstructed next to the second one, resulting in the reduction of the length of the chain, which upon further pulling eventually broke. Of course the particular details of the dynamical evolution might depend on the particular details of the initial conditions. However, the important conclusion that the formation of one-dimensional atomic chains depends on the existence of two-coordinated carbon dimers, as discussed above, holds true independently of the simulation details\cite{nec-suf}.

\begin{figure}[!ht]
\includegraphics[width=8.5cm]{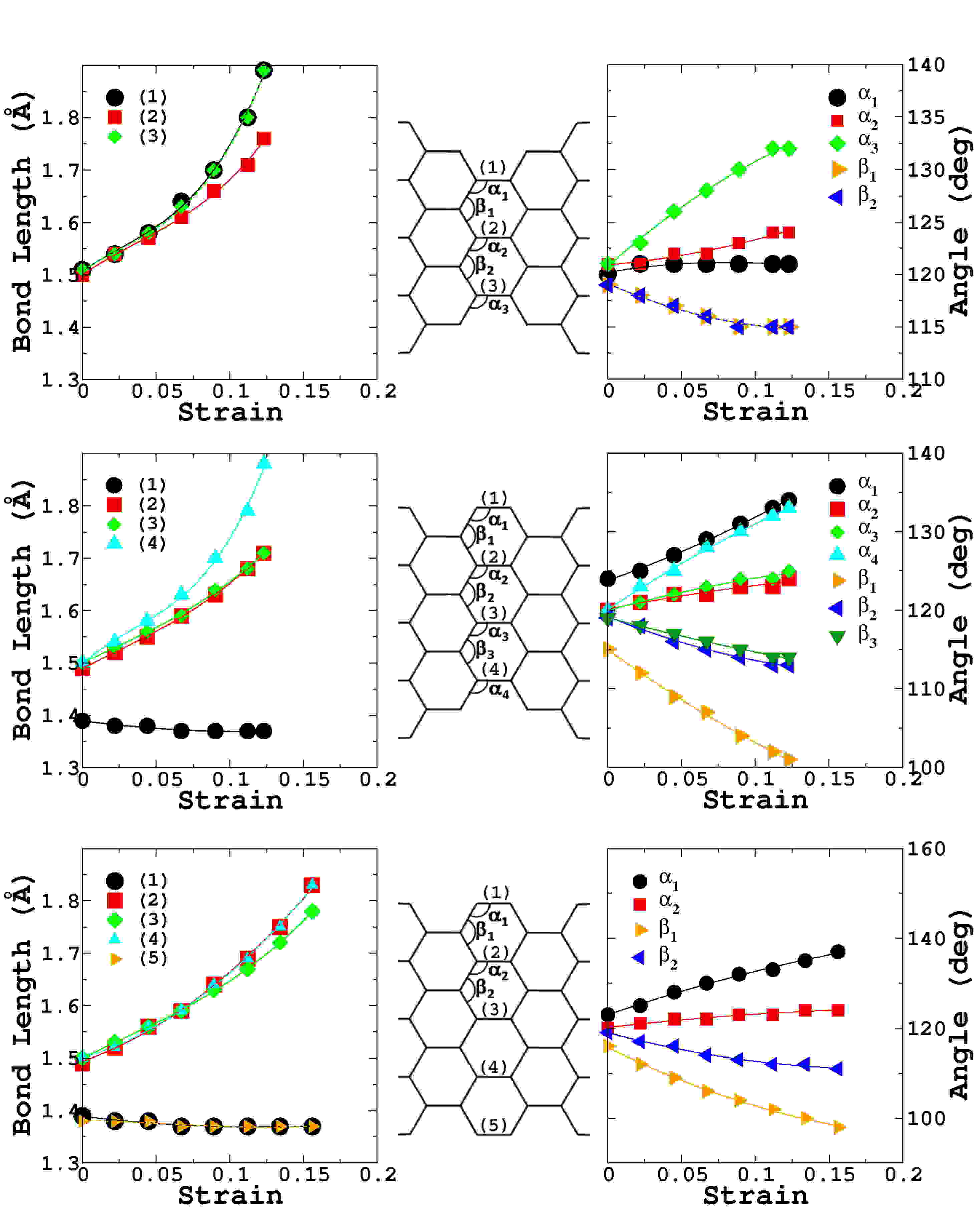}
\caption{Relevant bond lengths and angles (schematically shown in the central column panels), as a function of the strain, for the three systems studied (N1 - top panel; N2 - middle panel; N3 - bottom panel). We present the curves all the way up to the rupture of the first bond.}
\end{figure}

In order to understand why this is so, we analyze the average bond lengths and bond angles in the central neck region, as shown in Fig. 2. These averages were obtained during the time of MD simulation for each elongation, or strain. For the N1 configuration, the external bonds (bonds (1) and (3)) increase faster and they are where the rupture will occur. This is connected with the bond angle variations, where the external angle ($\alpha_3$) increases significantly. Now, for the N2 configuration, the bonds (2), (3) and (4) increase as the system is pulled, with bond (4) increasing much faster, which will lead to its rupture. The important point is that the bond length of the two-coordinated carbon dimer instead of increasing suffers actually a small decrease. Once again, these changes are correlated with the angular variations. There is a large decrease in the angle $\beta_1$ and an increase in the angle $\alpha_1$. This indicates that as the system is pulled the stress increase can be absorbed via angular variations for this two-coordinated carbon dimer, whereas the central neck bonds do not have the flexibility to do so because of their local constraints. Thus, all these internal neck bonds will break first, forming the one dimensional wire. For the N3 system a similar analysis holds true, as can be deduced from the values of the bond lengths and bond angles shown in Fig. 2.

\begin{figure}[!ht]
\includegraphics[width=8.5cm]{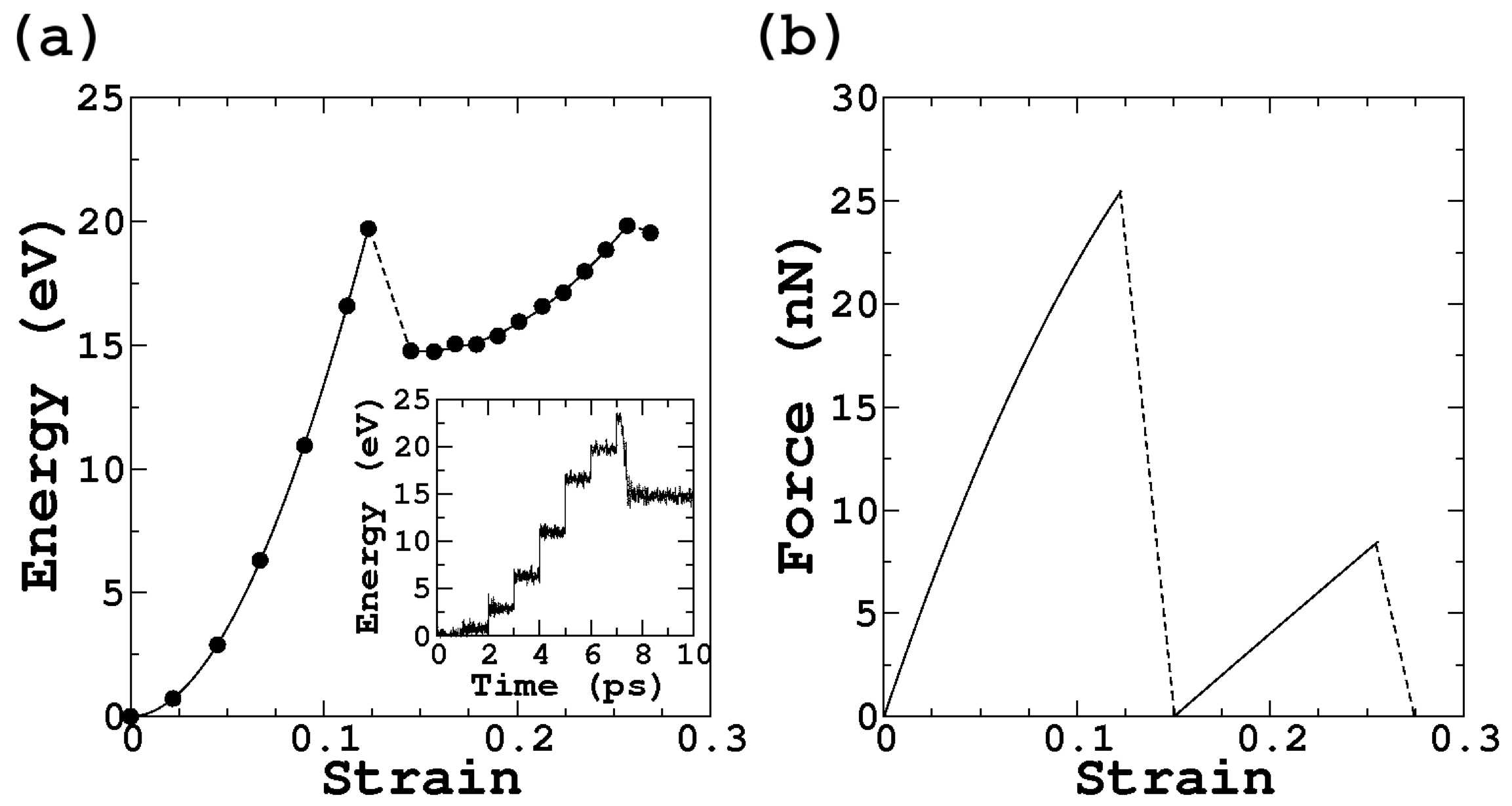}
\caption{(a) Average potential energy as a function of strain for the N2 system. Each one of the points corresponds to one of the steps shown in the inset. (b) External applied force required to cause a certain strain, obtained from Fig. 3(a).}
\end{figure}

In Fig. 3(a) we show the average potential energy (DFT energy) as a function of strain for the N2 system. Each one of the points in this graph were obtained by averaging over the MD simulation for a given length $L_W$ of the supercell, which corresponds to a certain strain (see the inset, where the average over each one of the steps corresponds to a point in the graph of energy versus strain). From this graph we can obtain the applied external force required to cause a certain strain, which is shown in Fig. 3(b).
One can see that the behavior is close to harmonic all the way up to the formation of the wire, at approximately 12\% strain. At this point the bonds break and the linear chain is formed. There is a relaxation of the system, with a sudden decrease of the potential energy. Further pulling of the system from this point again causes an elastic energy increase, basically absorbed at the one-dimensional carbon chain. When a force of approximately 8 nN is reached, there is a rupture of the chain. This number can be compared to the much smaller forces required to break gold atomic chains, which are of the order of 1.5 nN.

\begin{figure}[!ht]
\includegraphics[width=8.5cm]{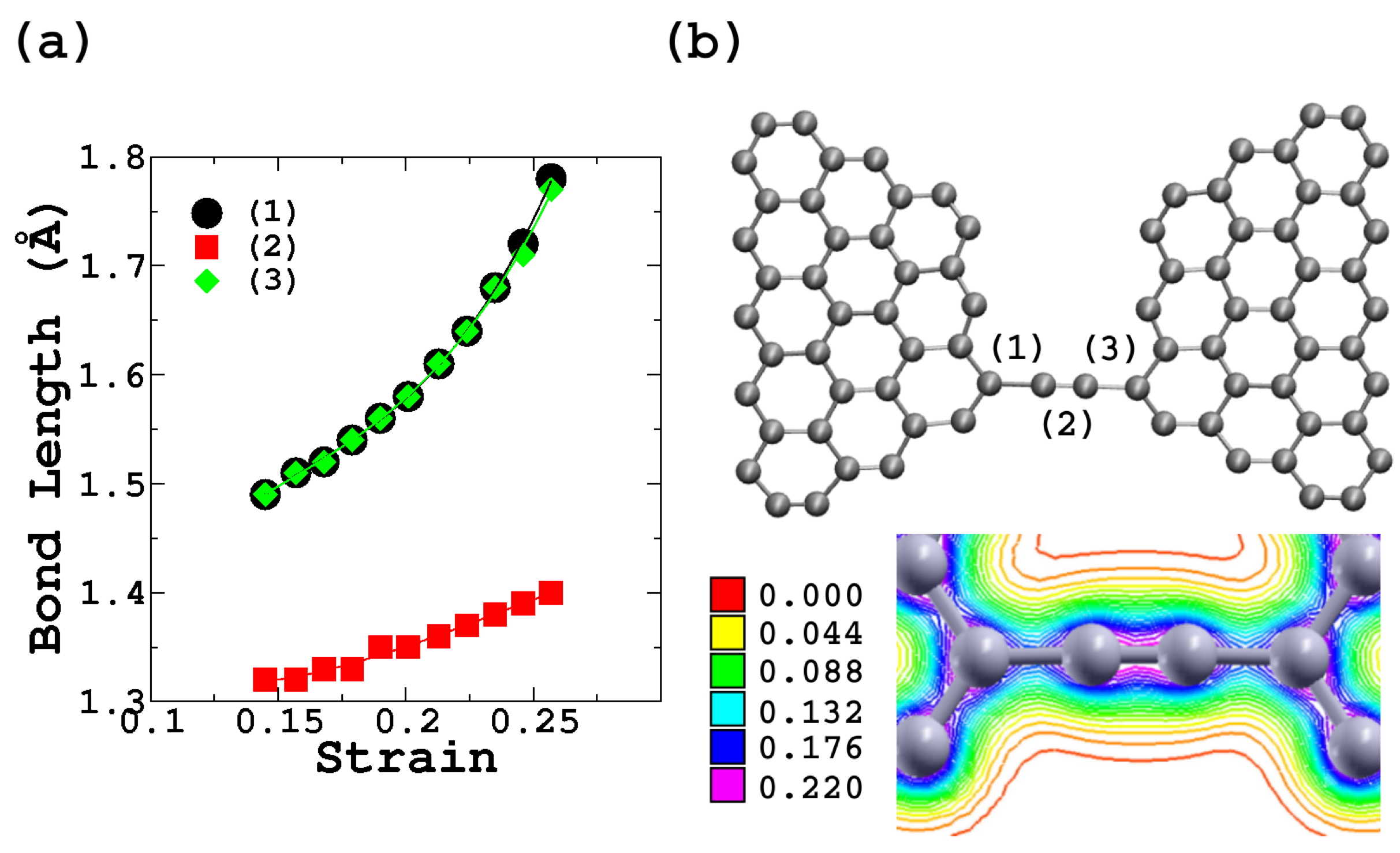}
\caption{Average interatomic distances between the suspended carbon atoms shown in the atomic model, for structure N2. In the bottom right panel a contour plot for the total electronic charge density is presented (units of $e$/\AA$^3$).}
\end{figure}

Regarding the bonding at the one-dimensional chain, we show in Fig. 4 the average interatomic distances between the suspended carbon atoms, for structure N2. As can be seen, there is an alternation between long and short distances, indicating the formation of single and triple bonds, which is a characteristic of polyynes\cite{poly}. The single bonds are the ones closer to the bulk. In Fig. 4 we also present contour plot for the total electronic charge density.
 The results confirm the assignment of single-triple-single bonds at the suspended chain.

\begin{figure}[!ht]
\includegraphics[width=8.5cm]{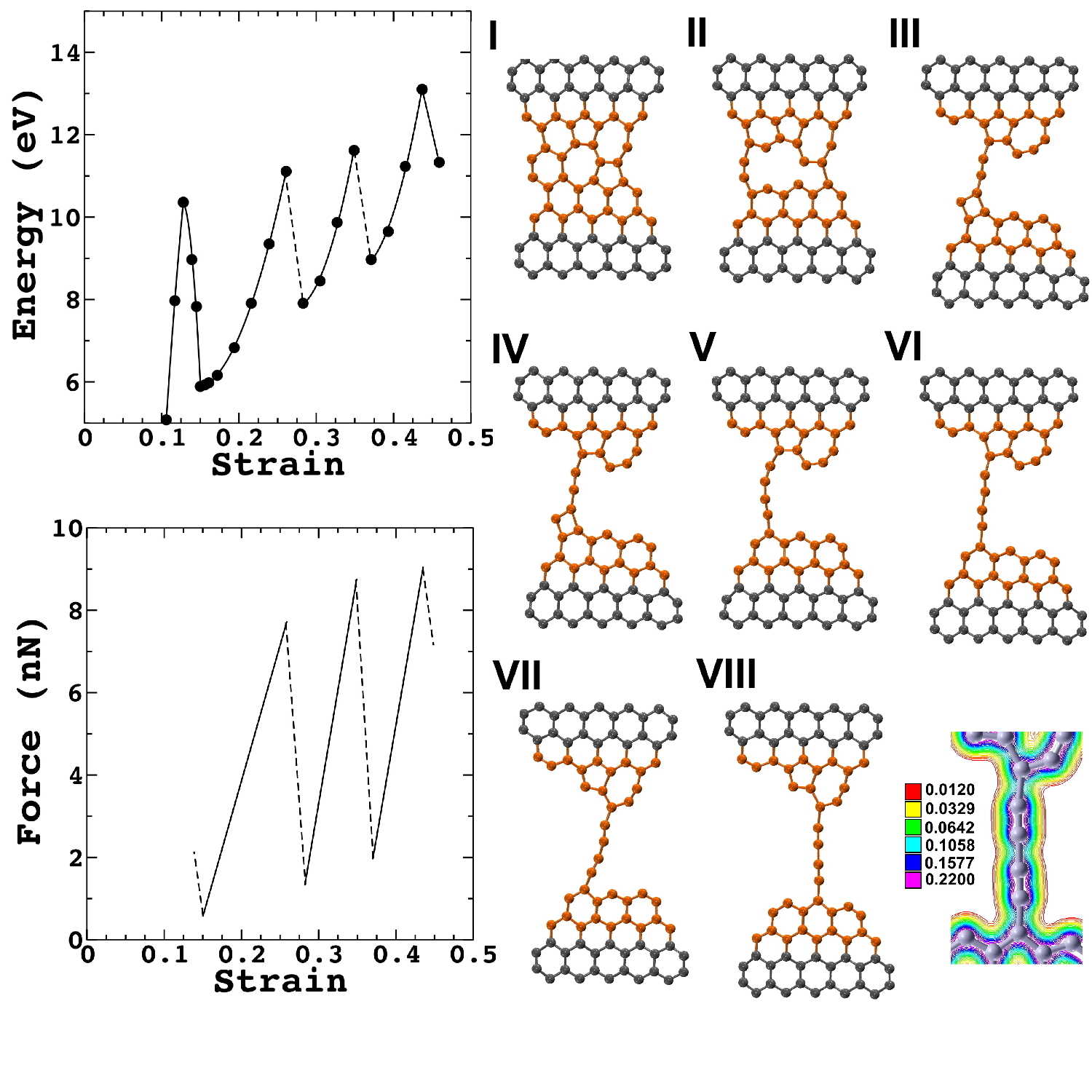}
\caption{Eight representatives geometries along the stretching process for a nanoribbon with a SW defect. The arrow in the last geometry marks the bond that breaks. The top left graph shows the variation of the average potential energy as a function of strain, where the numbers refer to the presented geometries, whereas the bottom left graph shows the external applied force required to cause a certain strain. In the bottom right panel a contour plot for the total electronic charge density is presented (units of $e$/\AA$^3$).}
\label{SW}
\end{figure}

We have also investigated the role of defects on the chain formation. Considering that the Stone-Wales (SW) is a possible defect to be formed under electron irradiation\cite{chuvilin}, we have performed a similar simulation as described above for the structure shown in Fig. \ref{SW}. This geometry still has the two-coordinated carbon dimer at one side of the neck, but has a SW defect at the other side. The SW defect can be seen as a rotation of one of the bonds of the hexagonal lattice, resulting in a local modification of four hexagons into two pentagons and two heptagons. In Fig. 5 we present eight geometries along the MD simulation (see also movie S4\cite{movies}). The corresponding average potential energy variation as a function of strain is shown in the top left panel. As can be seen, there are sequences of quasi-harmonic elastic energy increases followed by large atomic rearrangements. This behavior is similar to what has been observed in the evolution of suspended metallic nanowires\cite{agrait03,fios-nosso}. Initially there was the rupture of the bonds of the heptagon positioned in the middle of the ribbon (geometry II), followed by the total rupture at the opposite side of the two-coordinated carbon dimer (geometry III). The potential energy decreases from the geometry II to III due to the retraction of the system, which releases the accumulated elastic energy. This is the general mechanism related to the potential energy drops. This also resulted in a suspended one-dimensional carbon chain. However, as the simulation evolved, instead of simply observing the breaking of the chain, there was a rupture of one of the bonds of the hexagon where the chain was attached. This hexagon is the neighbor of the remaining pentagon of the SW defect. As a result, a larger suspended chain was formed (geometry V). Moreover, as the system was pulled the chain moved sideways, indicating that depending on the local geometry and bonding, the thermal energy might be sufficient to cause the motion of the wire (geometries VII-VIII). This migration was driven by the smaller spatial gap at the final connecting point, which leads to a smaller stress of the wire. Finally, further pulling causes the rupture of the chain, once more at one of the bonds attached to the bulk. In this case it was the bond connected to the remaining heptagon of the SW defect. For this longer chain we also observed an alternation between long and short bonds, indicating the formation of single and triple bonds, which is also supported by the contour plot of the total electronic charge density shown in the bottom right panel of Fig. 5.

In conclusion, we have analyzed the processes involved in the formation of linear carbon chains obtained by pulling graphene nanoribbons. We have shown that a crucial aspect necessary for the formation of these chains is the presence of dimers composed of two-coordinated carbon atoms at the edges of the ribbons at the neck position, otherwise there is simply the full rupture of the structure. 
The formation of longer chains may be associated with the presence of Stone-Wales defects close to the carbon dimers. Finally, in all situations where we there was the formation of suspended carbon chains, we have observed the presence of single and triple bonds, which is a characteristic of polyynes.

We acknowledge the financial support from the Brazilian Agencies FAPESP
(2005/59581-6; 2008/10503-1), CNPq and CAPES. EH and RBP have contributed equally to this work.


\begin{references}
\bibitem{hydrogen}
We have also performed simulations with the nanoribbon edges saturated with hydrogen, and the main conclusions of the paper are not altered.
\bibitem{nec-suf}
Note that this is a necessary but not sufficient condition. We have performed simulations for the N3 geometry, where the two-coordinated carbon dimers are present, but which due to the particular initial conditions evolved to a situation where the neck broke in one of the rows adjacent to the two-coordinated carbon dimers row, with a mechanism similar to the rupture of the N1 system.
\end{references}
\end{document}